\pdfoutput=1

\documentclass[superscriptaddress,amsmath,amssymb,twocolumn]{revtex4-1}

\usepackage{graphicx}
\usepackage{chemarr}
\usepackage{times,longtable}
\usepackage{hyperref}
\usepackage{color}
\usepackage{afterpage,mathtools}
\usepackage{soul,comment}
\usepackage[usenames,dvipsnames]{xcolor}
\usepackage{colortbl}
\usepackage{pbox}
\usepackage{capt-of}% or use the larger `caption` package
\usepackage{amsmath, amsthm}

\graphicspath{{large/}}

\bibpunct{[}{]}{,}{n}{}{,}

\newcommand{\red}[1]{\textcolor{black}{ #1}}

\newcommand{\grn}[1]{\textcolor{ForestGreen}{\bf #1}}

\newcommand{\inn}{\textrm{x}}
\newcommand{\out}{\textrm{g}}
\newcommand{\opt}{\textrm{opt}}

\newcommand{\avg}[1]{\left\langle #1 \right\rangle}
\newcommand{\avgs}[1]{\langle #1 \rangle}
\renewcommand{\bar}[1]{\overline{#1}}

\newcommand{\KL}{{\rm KL}}
\newcommand{\ovl}[1]{\overline{ #1}}
\renewcommand{\hat}[1]{\widehat{#1}}

\usepackage{chemarr}
\usepackage{epsf,mathtools}
\usepackage{graphicx}% Include figure files
\usepackage{dcolumn}% Align table columns on decimal point
\usepackage{bm}% bold math
\begin{document}

\title{Statistics of optimal information flow in ensembles of  regulatory motifs}

\author{Andrea Crisanti}
\affiliation{Dipartimento di Fisica, Sapienza Universit\`a di Roma, Rome, Italy}

\author{Andrea De Martino}
\affiliation{Soft and Living Matter Lab, Institute of Nanotechnology (CNR-NANOTEC), Consiglio Nazionale delle Ricerche, Rome, Italy}
\affiliation{Italian Institute for Genomic Medicine, Turin, Italy}

\author{Jonathan Fiorentino}
\affiliation{Dipartimento di Fisica, Sapienza Universit\`a di Roma, Rome, Italy}

\begin{abstract}
Genetic regulatory circuits universally cope with different sources of noise that limit their ability to coordinate   input and output signals. In many cases, optimal regulatory performance can be thought to correspond to configurations of variables and parameters that maximize the mutual information between inputs and outputs. Such optima have been well characterized in several biologically relevant cases over the past decade. Here we use methods of statistical field theory to calculate the statistics of the maximal mutual information (the `capacity') achievable by tuning the input variable only in an ensemble of regulatory motifs, such that a single \red{controller regulates} $N$ targets. Assuming (i) sufficiently large $N$, (ii) quenched random kinetic parameters, and (iii) small noise affecting the input-output channels, we can accurately reproduce numerical simulations both for the mean capacity and for the whole distribution. Our results provide insight into the inherent variability in effectiveness occurring in regulatory systems with heterogeneous kinetic parameters.
\end{abstract}

\maketitle
%\tableofcontents

\section{Introduction}

Regulatory processes in living cells are universally subject to noise. In many cases, it is essential that stochastic fluctuations affecting an upstream node of the regulatory network (e.g. a transcription factor, a cell-surface receptor, etc.) \red{are} not amplified as the biochemical cascade triggered by its activation (e.g. RNA transcription, a specific signaling pathway, etc.) proceeds to downstream nodes (e.g. proteins)\red{. Indeed,} efficient modulation of the cell's response in changing extracellular and/or endogenous conditions requires the output to be controllable with sufficient accuracy. In this light, noise processing appears to be a central task of regulatory circuits, and quantifying their noise-processing capability is an important theoretical question. 

During the past decade many studies have addressed this issue within an information theoretic framework in different contexts \cite{plos1,pnas1,pre1,pre2,pre3,pre4,pre7,pre8,bj,inforev,physbiol,pnas2,pre5,jpcm,jsp,pre6,ploscb,scirep}. The general idea behind this line of work is that optimal effectiveness of a regulatory module is achieved when the mutual information between input and output nodes is maximized. While individual  motifs may operate under non-trivial trade-offs in extended regulatory networks or even in populations of cells \cite{deeds}, optimal properties establish fundamental limits to noise processing by regulatory circuits. Therefore, their quantification allows in principle to isolate and characterize the physical ingredients that constrain information flow (e.g. noise sources, parameter configurations, etc.), leading to predictions that can be tested either in experiments (see e.g. \cite{pnas2,arxiv}) or via transcriptional or proteomic data analysis (see e.g. \cite{scirep}). 

Here we aim at extending the current theoretical picture by analyzing the statistics of optimal information flow in {\it ensembles} of regulatory motifs using tools of statistical field theory. \red{In specific, we consider ensembles generated by sampling the parameters characterizing input-output couplings from given probability distributions. This choice reflects a situation typical e.g. of sequence-specific couplings. In such cases, interaction parameters can be thought to change over time scales much longer (evolutionary) than those characterizing variations in molecular levels, and may therefore be considered  fixed (quenched) with respect to the faster variables. Generically speaking, as the number of components increases, ensemble properties often become less sensitive to the details of the interactions (i.e. the specific parameter values) while retaining a dependence on the parameter distribution \cite{nishi}. In this sense, they describe {\it typical} properties of systems of interacting units and provide a robust benchmark against which optimal properties can be gauged.} 

\red{For simplicity, we focus on the elementary case in which a single controller regulates a (possibly large) number of target nodes at stationarity \cite{pnas1}. We shall see that, in this class of systems, the existence of a well defined optimum allows for typical properties to get closer and closer to it as the number of targets increases. This suggests that, at least to some degree, sufficiently large and centralized regulatory elements might be relatively insensitive to interaction details. Biological instances in which such a situation could be realized are found  e.g. in miRNA-mediated post-transcriptional regulatory networks (where some miRNAs are known to control the expression of a large number of target RNAs \cite{raj}), in the coupling of transcription factors to promoters in bacteria \cite{wolf,gerland} and in certain emergent properties of  metabolic networks \cite{vn,cafba}.} However, for sakes of brevity and to focus the article on our main goal, we shall not address extensively the biological context underlying these models, referring the reader e.g. to \cite{jpcm} and \cite{scirep} for detailed discussions of the cases of transcriptional and post-transcriptional regulation, respectively. 

After briefly recapping the background and stating the problem and our strategy (Sections II and III), we concentrate on the ensemble of optimal motifs generated by a probability distribution of (the values of) kinetic parameters. Specifically, we first compute the average capacity, i.e. the mean value of the maximum mutual information exchanged between \red{controller} and targets (Sec. IV.A), and then derive an expression for the probability distribution of the capacity (Sec. IV.B). Analytical results are compared against numerical experiments in Section V. Finally, in Sec. VI we provide an outlook of our results.

\begin{figure*}
	\includegraphics[width=0.9\textwidth]{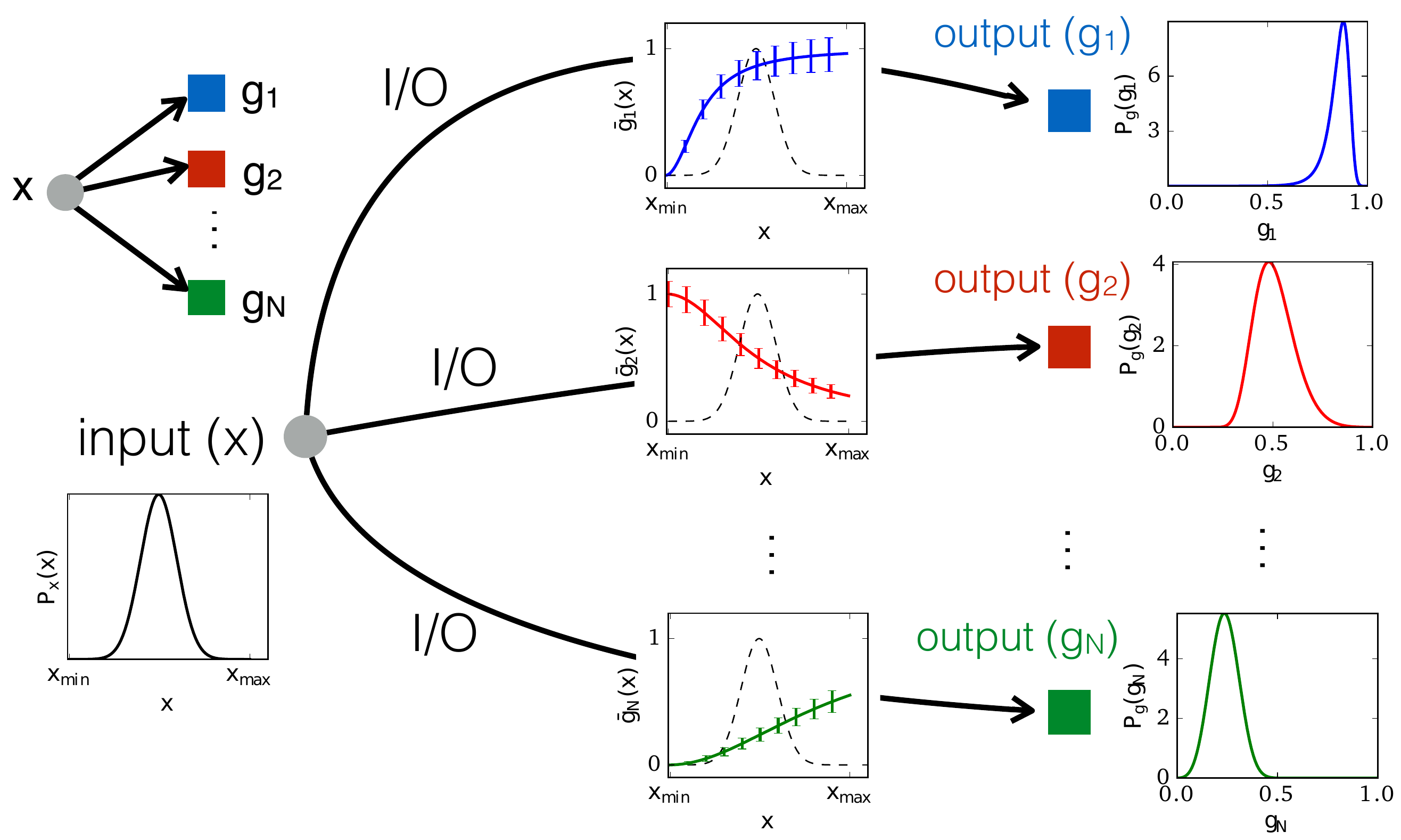}
	\caption{Scheme of one controller (e.g. a transcription factor) regulating $N$ independent targets. $P_\inn(x)$ is the probability distribution of the controller level $x$, \red{recalled by the dashed black lines in middle panels}. The channel is represented by the mean target levels $\{{\bar{g}}_i(x)\}_{i=1}^N$ \red{(solid lines in middle panels)}, modeled as Hill functions of $x$ (the controller can either activate or repress each of the $N$ targets). The error bars represent $\pm \sigma_i(x)$, where $\sigma_i^2(x)$ is the intrinsic noise variance of the i-th target. The presence of noise induces a probabilistic relationship between the levels of the controller and the targets: the outputs of the regulatory network are the probability distributions $\{P_{\rm g}(g_i)\}_{i=1}^N$, whose shape depends on the matching between the controller distribution and the noisy channel.}
	\label{fig:FIG1}
\end{figure*}

\section{Background}

%\subsection{Basic framework}
%\label{sec:basic}

We follow \cite{pre2} and consider a single \red{input} node, representing e.g. a transcription factor or a regulatory RNA,  controlling $N$ targets, see Fig. \ref{fig:FIG1}. We assume that targets interact through the controller exclusively (i.e. there is no direct coupling between targets). The state of each node is described by a concentration variable. We let $x$ denote the controller level, which we measure in units of a pre-assigned concentration value, and assume it ranges from $x_{\min}$ to $x_{\max}$. The vector $\boldsymbol{g}=\{g_i\}$ will instead represent target levels, with $i=1,\ldots,N$. For sakes of simplicity, we assume that $g_i$'s are dimensionless and that $g_i\in[0,1]$ for each $i$, amounting to a re-scaling of each $g_i$ by its maximum attainable level.

\begin{figure}
	\includegraphics[width=0.48\textwidth]{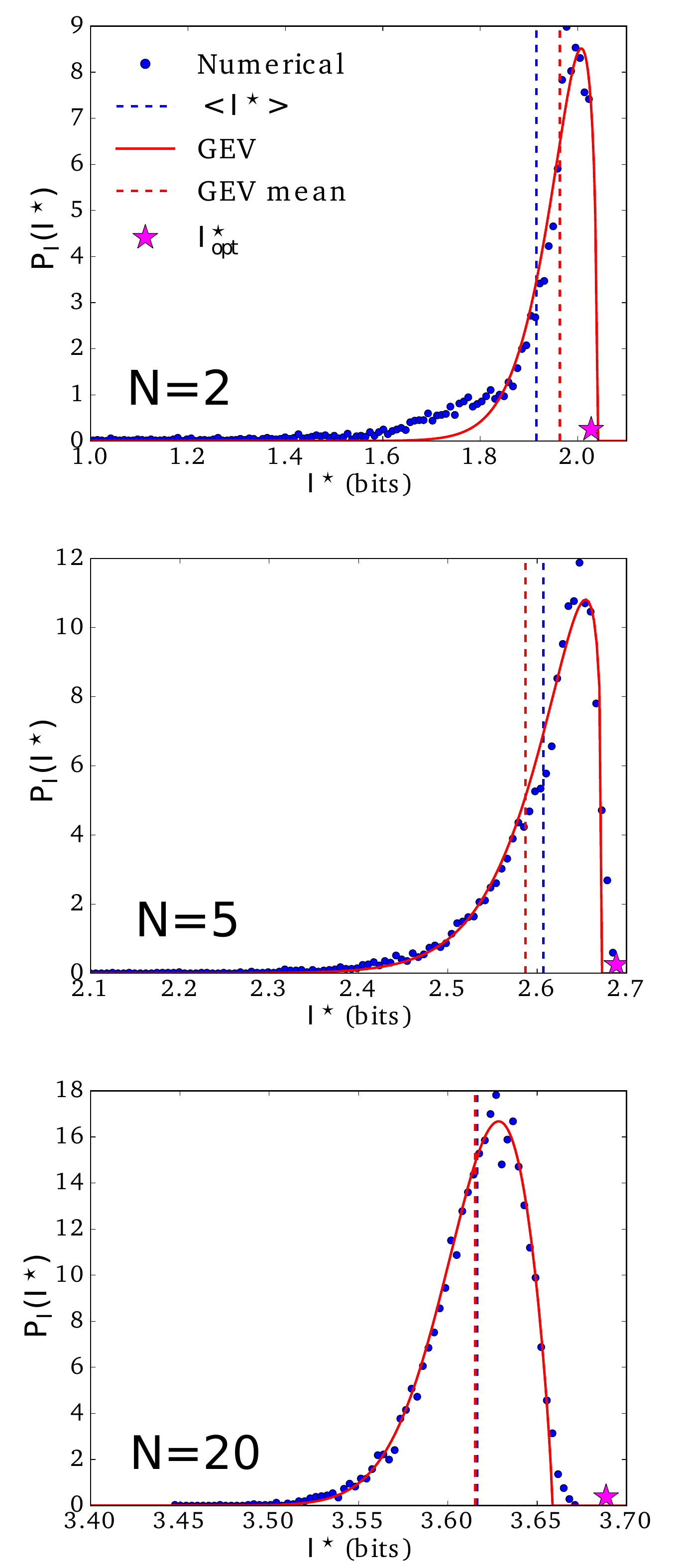}
	\caption{Blue markers: probability distributions $P_{\rm I}(I^\star)$ obtained numerically for ensembles of $10^4$ regulatory motifs with different $N$ constructed by choosing $\bar g_i$ and $\sigma_i^2$ as in (\ref{gbar}) and (\ref{sigma2}), by sampling each $K_i$ independently from a uniform distribution in $[x_{\min},x_{\max}]$, with $x_{\min}=10^{-2}$, $x_{\max}=1$, and by setting $h_i=2$ for each $i=1,\dots,N$. The red line represents the best fit to a GEV distribution \red{\cite{gev}}. Blue and red vertical lines mark, respectively, the positions of the numerical mean and of the mean of the GEV fitted to the data. The purple star indicates $I^\star_\opt$, Eq. \eqref{opt}. The maximum number of \red{target} molecules was set to $M_{\max}=100$.}
\label{fig:FIG2}
\end{figure}

In a purely static setting, each $g_i$ is taken to be determined independently from $x$ via a regulatory channel described by the conditional probability density $P(g_i|x)$, which encodes for the complex physical processes that map the input variable $x$ into the output variables $g_i$. Again following \cite{pre2}, $P$ is taken to be Gaussian with mean $\bar g_i(x)$ and variance $\sigma_i^2(x)$, i.e.
\begin{equation}\label{channel}
P(g_i|x)=\frac{1}{\sqrt{2\pi\sigma_i^2(x)}}\exp\left[-\frac{(g_i-\bar g_i(x))^2}{2\sigma_i^2(x)}\right]~~.
\end{equation}

In short, one can imagine that the regulatory motif responds stochastically to the value of $x$ being fed into it, with an average response given by $\bar g_i$ and fluctuations described by $\sigma_i^2$. Whether $x$ tends to activate or repress the $i$-th target is determined by the behaviour of $\bar g_i$ (and more complicated dependencies can be realized by modulating the topology \cite{pre3,pre4}), whereas $\sigma_i^2$ accounts for the different sources of noise that contribute to the overall stochasticity of the channel $x\to g_i$. Detailed discussions of the biological ingredients of both $\bar g_i$ and $\sigma_i^2$ can be found e.g in \cite{pnas0,prl,plos2}. For instance, if the controller is a transcription factor (TF) that activates the transcription of the targets' RNAs upon binding to the DNA, a minimal, biologically plausible model for $\sigma_i^2$ and $\bar g_i$ is given by
\begin{gather}\label{gbar}
\bar g_i(x)=\frac{x^{h_i}}{x^{h_i}+K_i^{h_i}}~~,\\
\label{sigma2}
\sigma_i^2(x)=\frac{1}{M_{\max}}\left[\bar g_i(x)+x\left(\frac{\partial\bar g_i}{\partial x}\right)^2\right]~~,
\end{gather}
where $M_{\max}$ stands for the maximum achievable number of \red{target} molecules. In short, the Hill function (\ref{gbar}) generically describes the increase of $\bar g_i$ with $x$ that is expected when the controller enhances the synthesis of the target. The dissociation constant $K_i$ gives the value of $x$ for which $\bar g_i$ attains half of its maximum, while $h_i>0$ (the Hill index) quantifies the steepness of $\bar g_i$ for values of $x$ around $K_i$ (when $\bar g_i$ is most sensitive to changes in $x$, as $\bar g_i\ll 1$ for $x\ll K_i$ and $\bar g_i\simeq 1$ for $x\gg K_i$). Equation (\ref{sigma2}) instead details the contributions of standard molecular noise (first term) and of the randomness associated to diffusion-mediated DNA-TF interactions (second term) to the stochasticity of $g_i$ at any given $x$. Note that the specific forms of $\bar g_i$ and $\sigma_i^2$ are immaterial for our theory. 

Generically, if $x$ is sampled from an ensemble of values described by the  probability density $P_\inn(x)$, the quantity
\begin{gather}
P_\out(\boldsymbol{g})=\int_{x_{\min}}^{x_{\max}}P(\boldsymbol{g}|x)P_\inn(x)dx
\end{gather}
with $P(\boldsymbol{g}|x)=\prod_{i=1}^N P(g_i|x)$ describes the statistics of the output vector $\boldsymbol{g}$. In order to quantify the degree of control over $\boldsymbol{g}$ that can be exerted through $x$, one can employ the mutual information (in bits) between the input and output variables, namely
\begin{multline}
I(x;\boldsymbol{g})=\int d\boldsymbol{g} \,dx\, P(x,\boldsymbol{g})\log_2\frac{P(x,\boldsymbol{g})}{P_\inn(x)P_\out(\boldsymbol{g})}\\
=\int_{x_{\min}}^{x_{\max}}dx\,P_\inn(x)\int d\boldsymbol{g} \,P(\boldsymbol{g}|x)\log_2\frac{P(\boldsymbol{g}|x)}{P_\out(\boldsymbol{g})}~~,
\end{multline}
where $P(x,\boldsymbol{g})=P(\boldsymbol{g}|x)P_\inn(x)$ stands for the joint probability density of $x$ and $\boldsymbol{g}$. In rough terms, the number of different ``states'' for the vector $\boldsymbol{g}$ that can be reliably distinguished based on the noisy input variable $x$ is approximately given by $ 2^{I(x;\boldsymbol{g})}$. In a transcriptional regulatory setting, this would correspond to the number of distinct ``expression profiles'' of the $N$ targets that might be obtained by tuning the level of the controller. Hence larger values of $I$ can generically be associated to more refined degrees of control over the output layer. In particular, for a fixed input/output channel $P(\boldsymbol{g}|x)$, one can probe the limits to information flow by searching for the input distribution $P_\inn$ that maximizes $I$. 

When $\sigma_i^2(x)$ is sufficiently small for all $x$ (``small noise approximation''), the maximization problem can be solved analytically \cite{pre2}. The mutual information takes the form (see Appendix A for a short recapitulation of this scenario)
\begin{multline}\label{ISNA}
I(x;\mathbf{g})=S[P_\inn]+\\
+\int_{x_{\min}}^{x_{\max}}dx\,P_\inn(x)\log_2\sqrt{\frac{1}{2\pi e}\sum_{i=1}^N\frac{1}{\sigma_i^2(x)}\left(\frac{\partial\bar g_i}{\partial x}\right)^2}~~,
\end{multline}
where
\begin{equation}
S[P_\inn]=-\int_{x_{\min}}^{x_{\max}}dx\,P_\inn(x)\log_2 P_\inn(x)
\end{equation}
is the entropy of $P_\inn$. By variational differentiation of the above expression over $P_\inn$ (with a constraint enforcing normalization) one finds that the input distribution 
\begin{gather}
P^\star_\inn(x)=\frac{1}{Z}\left[\frac{1}{2\pi e}\sum_{i=1}^N\frac{1}{\sigma_i^2(x)}\left(\frac{\partial\bar g_i}{\partial x}\right)^2\right]^{1/2}~~,\label{Pstar}\\
Z=\int_{x_{\min}}^{x_{\max}} \left[\frac{1}{2\pi e}\sum_{i=1}^N\frac{1}{\sigma_i^2(x)}\left(\frac{\partial\bar g_i}{\partial x}\right)^2\right]^{1/2}dx\label{zeta}~~,
\end{gather}
exploits the channel described by (\ref{channel}) optimally, i.e. maximizes the mutual information between $x$ and $\boldsymbol{g}$. For $P_\inn=P_\inn^\star$, in particular, one gets
\begin{equation}\label{Istar}
I(x;\boldsymbol{g})\equiv I^\star=\log_2 Z~~.
\end{equation}
In information-theoretic terms, the quantity $I^\star$ represents the {\it capacity} of the given input/output channel.

\section{Problem statement and calculation strategy}

In concrete cases, the value of $I^\star$ depends on the specifics of the functions $\sigma_i^2$ and $\bar g_i$. For fixed choices of the above functions, e.g. as in \eqref{gbar} and \eqref{sigma2}, an important reference value is given by the maximum of $I^\star$ over the parameter space, i.e. 
\begin{equation}\label{opt}
I^\star_{\rm opt}=\max_{\boldsymbol{K,h}} \, I^\star~~,
\end{equation} 
where $\boldsymbol{K}=\{K_i\}$ and $\boldsymbol{h}=\{h_i\}$, the parameters characterizing \eqref{gbar} and \eqref{sigma2}, are assumed to take on values in prescribed ranges. Quantities like (\ref{opt}) have been studied extensively in the literature (see e.g. \cite{pre2,pre3,pre4}). On the other hand, one may be interested in the statistics of the capacity $I^\star$ for an {\it ensemble} of regulatory motifs defined, for any given choice of $\sigma_i^2$ and $\bar g_i$, by a probability distribution $Q$ for the parameter vectors. Numerical results obtained for small $N$ and in the small noise limit have shown that, at least in some cases, the optimum of $I^\star$ in the parameter space can be rather broad, implying a relatively weak dependence of the capacity on parameters \cite{pre2,ploscb,scirep}. It is therefore important to quantify ensemble properties more precisely.

The probability distribution of $I^\star$ (we shall denote it by $P_{\rm I}(I^\star)$) induced by a probability distribution of parameters is defined over the interval $[0,I^\star_{\rm opt}]$ and can be easily computed numerically from (\ref{zeta}) and (\ref{Istar}). Fig. \ref{fig:FIG2} shows results obtained by choosing $\bar g_i$ and $\sigma_i^2$ as in (\ref{gbar}) and (\ref{sigma2}), together with the best fits to a Generalized Extreme Value (GEV) distribution \red{\cite{gev}}. Notice how the mean lies within a few \red{per cent} of the optimum in each case (and moves closer to it as $N$ increases).

Here we aim at studying the distribution of $I^\star=\log_2 Z$ and its mean value $\avg{I^\star}$ using statistical mechanics tools. Our starting point will be Eq. (\ref{zeta}), which we re-write as
\begin{equation}\label{zstar2}
Z=\sqrt{N}\int_{x_{\min}}^{x_{\max}}dx\sqrt{\lambda(x)}
\end{equation}
upon defining 
\begin{gather}\label{lambda}
\lambda(x)=\frac{1}{N}\sum_{i=1}^N f_i(x)~~,\\
f_i(x)=\frac{1}{2\pi e \sigma_i^2(x)}\left(\frac{\partial\bar g_i}{\partial x}\right)^2\label{accai}~~.
\end{gather}
Randomness in the parameters of the functions $\bar{g}_i(x)$ and of $\sigma_i^2(x)$, and hence in the $f_i(x)$, makes $Z$ a random variable. In turn, its probability density depends on the probability distribution of parameters via the function $\lambda(x)$ which expresses the arithmetic mean of the $f_i$s at fixed $x$. It is reasonable to think that, if the cumulants of the $f_i$s do not grow too fast with $N$, the probability density functional ${\cal P}\bigl[\lambda]$ of  $\lambda(x)$ will be approximately Gaussian for large $N$, i.e.
\begin{widetext}
\begin{equation}
\label{eq:P_lam_gau}
 {\cal P}\bigl[\lambda] \simeq \frac{1}{{\cal N}} \,
     \exp\left[-\frac{N}{2} \int dx\, dx'\,\bigl[\lambda(x) - \mu(x)\bigr]\, \Delta^{-1}(x,x')\,
              \bigl[\lambda(x') - \mu(x')\bigr]
                                       \right]~~,
                                       \qquad N\gg 1,
\end{equation}
\end{widetext}
where ${\cal N}$ is a normalization factor, $\mu(x)$ denotes the mean of $\lambda(x)$, namely
\begin{equation}
\label{eq:mu_1p}
  \mu(x) = \frac{1}{N}\sum_{i=1}^{N}\langle f_i(x)\rangle_c~~,
\end{equation}
and $\Delta^{-1}(x,x')$ stands for the inverse of
\begin{equation}
\label{eq:Delta_2p}
  \Delta(x,x') = \frac{1}{N}\sum_{i,j=1}^{N}\langle f_i(x)\, f_j(x')\rangle_c~~.
\end{equation}
In the above formulas, averages are taken over the probability distribution functional of $f_i(x)$ or, equivalently, over parameters (with probability distribution $Q$), while the subscript ``$c$'' denotes cumulants or connected correlation functions, depending on the context. Note that, if $f_i(x)$ and $f_j(x')$ for $i\not=j$ are statistically independent, only terms with $i=j$ survive in expression (\ref{eq:Delta_2p}), which simplifies to
\begin{equation}
  \Delta(x,x') =  \frac{1}{N}\sum_{i=1}^{N}\langle f_i(x)\, f_i(x')  \rangle_c~~.
\end{equation}

Eq. (\ref{eq:P_lam_gau}) suggests that, for large enough $N$, means of quantities involving $Z$ may be evaluated by relying on the first two cumulants of $f_i(x)$ only. Moreover, if the (quenched) average capacity 
\begin{equation}\label{que}
\avg{I^\star}=\avg{\log_2 Z}
\end{equation}
can be approximated by the annealed estimate 
\begin{equation}\label{ann}
\avg{I^\star}_{\rm ann}=\log_2\avg{Z}~~,
\end{equation}
only cumulants of $f_i(x)$ at {\it fixed} value of $x$ are required (see (\ref{zstar2})). The error made upon replacing \eqref{que} with \eqref{ann}, an estimate of which is given in Appendix B, numerically turns out to be extremely small already for $N=5$. Therefore, we shall begin by computing $\avg{I^\star}_{\rm ann}$ analytically in the Gaussian approximation, showing that numerical results for \eqref{que} are remarkably well reproduced by our analytical result for \eqref{ann} even for small $N$. Next, we shall obtain an expression for $P_{\rm I}$ via a direct computation starting from (\ref{zstar2}), still within the Gaussian scenario. A comparison with numerical results will again turn out to be very good, although the accuracy will now depend more strongly on $N$. 

Improvements to the Gaussian approximation are not hard to include, at least formally. To illustrate how, we shall consider the lowest-order correction, leading to a cubic theory based on the third-order cumulant
\begin{equation}
\label{eq:cubic_v}
  g_3(x,x',x'') = \frac{1}{N}\sum_{i,j,k=1}^{N}\langle f_i(x)\, f_j(x')\, f_k(x'')\rangle_c~~.
\end{equation}
The possibility of keeping only diagonal terms in the above expression (i.e. with $x=x'=x''$) greatly simplifies calculations. \red{We will see that cubic corrections provide a better description of the statistics of $\lambda(x)$ for small $x$, especially when $x_{\max}\gg x_{\min}$, and improve the Gaussian estimate for $\avg{I^\star}$.}%While not especially important to describe $\avg{I^\star}$, cubic corrections provide a better description of the statistics of $\lambda(x)$ for small $x$, especially when $x_{\max}\gg x_{\min}$. 

Two important remarks are in order. First, since the comparison with numerical (or, possibly, experimental) data becomes harder as the order of the cumulants increases (even for statistically independent $f_i(x)$), the predictions obtained with the Gaussian theory provide, in our view, a useful benchmark {\it per se}. Secondly, we shall see that our approach does not require to prescribe the details of the probability distribution  $Q$ of parameters. Likewise, the particular choices of $\bar g_i$ and $\sigma_i^2$ are immaterial for the calculation. We shall therefore only specify them upon comparing with numerical results. To fix ideas, though, we shall henceforth indicate the parameters entering the definitions of $\bar g_i$ and $\sigma_i^2$ by $\boldsymbol{K}$ and $\boldsymbol{h}$, as in \eqref{gbar} and \eqref{sigma2}, and their probability distribution by $Q(\boldsymbol{K},\boldsymbol{h})$.

\section{Analytical results}

\subsection{Average channel capacity}
\label{sub:ann}

As said above, we shall approximate the average capacity by  the quantity
\begin{equation}\label{annealed}
\avg{I^\star}_{{\rm ann}}=\log_2\avg{Z}~~,
\end{equation}
where the average $\avg{\cdots}$ is over the parameters $\boldsymbol{K}$ and $\boldsymbol{h}$, i.e.
\begin{equation}\label{avg1}
  \avg{(\cdots)}=\int d\boldsymbol{K}d\boldsymbol{h}\,Q(\boldsymbol{K},\boldsymbol{h})\, (\cdots)~~,
\end{equation}
or, equivalently, over the functions $f_i(x)$, i.e.
\begin{equation}\label{avg2}
   \avg{(\cdots)}=\int {\cal D}\boldsymbol{f}\, {\cal P}[\boldsymbol{f}]\,  (\cdots)~~,
\end{equation}
with probability distribution functional
\begin{multline}
  {\cal P}[\boldsymbol{f}] = \int dx \int d\boldsymbol{K}d\boldsymbol{h} \,Q(\boldsymbol{K},\boldsymbol{h})\times\\
      \times\prod_{i=1}^{N} \delta\Bigg[f_i(x) - \frac{1}{2\pi e \sigma_i^2(x)}\left(\frac{\partial\bar g_i}{\partial x}\right)^2\Bigg]~~.
  \end{multline}
While (\ref{avg1}) and (\ref{avg2}) are equivalent, the use of one vs. the other may depend on the context. 

Using \eqref{zstar2}, the quantity $\avg{Z}$ can be written as
\begin{equation}\label{zstar3}
\avg{Z}=\sqrt{N}\int_{x_{\min}}^{x_{\max}}dx \int_0^\infty d\lambda\, P_\lambda(\lambda;x)\, \sqrt{\lambda}~~,
\end{equation}
where
\begin{equation}
P_{\lambda}(\lambda;x)=\int d\boldsymbol{f}\, P_\textrm{f}(\boldsymbol{f},x)\, 
                  \delta\left(\lambda-\frac{1}{N}\sum_{i=1}^N f_i\right)~~,
\end{equation}
and 
$P_\textrm{f}(\boldsymbol{f};x) = \left\langle \prod_{i=1}^{N} \delta\left(f_i(x) - f_i\right)\right\rangle$ is the probability density of the value of the functions $f_i(x)$ at the given $x$.
Using the integral representation of the Dirac $\delta$-function, one gets
\begin{multline}
    P_\lambda(\lambda;x)= N \int_{-\infty}^{+\infty} \frac{d\phi}{2\pi}\, 
                   \int d\boldsymbol{f}\, P_\textrm{f}(\boldsymbol{f};x)\, e^{-i\phi\left(N\lambda-\sum_i f_i\right)}=\\
             =N  \int_{-\infty}^{+\infty} \frac{d\phi}{2\pi}\,
                   e^{-Ni\phi\lambda+\ln \bigl\langle e^{i\phi\sum_i f_i}\bigr\rangle_x}~~.
\end{multline}
The subscript `$x$' here indicates that the average is taken at fixed $x$. The second term in the exponent can be expressed using the cumulants of the quantity $\sum_i f_i$ as
\begin{equation}
 \ln \bigl\langle e^{i\phi\sum_i f_i}\bigr\rangle_x = 
             \sum_{n \geq 1}\frac{1}{n!} (i\phi)^n \Bigl\langle \Bigl(\sum_i f_i\Bigl)^n\Bigr\rangle_{x,c}~~.
\end{equation}
Thus, defining
\begin{align}
 \label{eq:mu_x}
 \mu(x) &= \frac{1}{N}\sum_{i=1}^N \langle f_i\rangle_x~~, \\
 \label{eq:Delta_x}
 \Delta(x) &= \frac{1}{N}\sum_{i,j=1}^N \langle f_i f_j\rangle_{x,c}~~, \\
  \label{eq:gn_x}
 g_n(x) &= \frac{1}{N}\sum_{i_1,\dotsc,i_n=1}^N \langle f_{i_1} \dotso f_{i_n}\rangle_{x,c}~~, 
 \quad (n\geq 3)~~,
\end{align}
and introducing the {\it action}
\begin{equation}
\label{eq:S_x}
  S(\phi;x) = \frac{\Delta(x)}{2} \phi^2 - \sum_{n\ge 3} \frac{g_n(x)}{n!} (i\phi)^n~~,
\end{equation}
the function $P_\lambda(\lambda;x)$ can be re-cast as
\begin{equation}
\label{eqn:P_lambda1}
\begin{split}
   P_\lambda(\lambda;x) &=N \int_{-\infty}^{+\infty} \frac{d\phi}{2\pi}\,
                                  e^{-N\bigl[S(\phi;x) - i\phi(\mu(x) -\lambda)\bigr]} \\
                         &= \frac{N {\cal A}}{2\pi} \left. e^{NW(J;x)}\right|_{J= i(\mu(x) - \lambda)}~~,
\end{split}                          
\end{equation}
where ${\cal A}\equiv{\cal A}(x) = \int d\phi\, e^{-NS(\phi;x)}$ and $W(J;x)$ is the cumulant generating function of the random variable $\phi$, whose probability density reads $e^{-NS(\phi;x)}/{\cal A}(x)$.
The normalization constant ${\cal A}(x)$ ensures that $W(J=0;x)=0$.

We stress that the cumulants \eqref{eq:mu_x}, \eqref{eq:Delta_x} and \eqref{eq:gn_x} depend on $x$. Moreover, \eqref{eq:mu_x} coincides with \eqref{eq:mu_1p}, while \eqref{eq:Delta_x} corresponds to $\Delta(x,x)$ as defined in \eqref{eq:Delta_2p}. Similarly, the quantity $g_n(x)$ defined in \eqref{eq:gn_x} is simply $g_n(x_1,\dotsc,x_n)$ with $x_1=\dotso=x_n = x$ as defined in \eqref{eq:cubic_v} for $n=3$. For simplicity, we shall henceforth retain the explicit dependence on $x$ only in the notation $P_\lambda(\lambda;x)$, dropping it elsewhere.

It is also worth noting that $\langle \lambda \rangle = \mu$, $\langle \lambda^2 \rangle_c = \Delta/N$ and $\langle \lambda^n \rangle_c = g_n/N^{n-1}$ for $n\geq 3$. Thus, if $\mu$, $\Delta$ and $g_n$ are all of $O(1)$ as $N\gg 1$, the cumulants of $\lambda$ vanish more rapidly with $N$ as the order increases. It follows that $P_\lambda(\lambda;x)$ can be well approximated, for large $N$, by a Gaussian with mean $\mu$ and variance $\Delta/N$. This result could have been anticipated from the Central Limit Theorem. The procedure just discussed however allows for a systematic treatment of corrections to this picture. Indeed, in the following we shall first compute $P_\lambda(\lambda;x)$ in the Gaussian scenario by retaining only the first (quadratic) term in $S(\phi)$, and then consider the first correction to it, obtained by including the term proportional to $g_3$ in \eqref{eq:S_x}. In each case, the quantity $\avg{I^\star}_{\rm ann}$ will ultimately be obtained from \eqref{annealed} and \eqref{zstar3} using the expressions for $P_\lambda(\lambda;x)$ derived in each case.

\subsubsection{Gaussian approximation (Central Limit Theorem)}

Neglecting all terms but the quadratic in \eqref{eq:S_x} one easily arrives at the Gaussian form
\begin{equation}
\label{eqn:P_lambda_2}
\begin{split}
   P_\lambda(\lambda;x) &=N \int_{-\infty}^{+\infty} \frac{d\phi}{2\pi}\,
                                  e^{-N\bigl[\frac{1}{2}\Delta \phi^2 - i\phi(\mu -\lambda)\bigr]} \\
                         &= \sqrt{\frac{N}{2\pi\Delta}}\, e^{-\frac{N}{2\Delta}(\lambda -\mu)^2}~~,
\end{split}                          
\end{equation}
from which $\avg{\lambda} = \mu$  and $\avg{\lambda^2}_c =\Delta/ N$ readily follows.

\subsubsection{Cubic approximation: leading order}

The first correction to the Gaussian approximation is obtained by including the third order term in the action $S(\phi)$, Eq. \eqref{eq:S_x}, which now reads
\begin{equation}
\label{eq:S_3}
S(\phi)= \frac{1}{2} \Delta\phi^2+i \frac{g_3}{3!}\phi^3~~.
\end{equation}
The simplest way of computing the corrections induced by the third order term is by inserting \eqref{eq:S_3} into the first line of \eqref{eqn:P_lambda1} and expanding the resulting expression in powers of $\Delta/N$. A better approximation consists in using the second line of \eqref{eqn:P_lambda1} and computing the cumulant generating function $W(J)$ of the theory governed by $S(\phi)$ from its Legendre Transform  $\Gamma(\varphi)$ (also known as `effective potential'), defined via
\begin{equation}
\label{eqn:def_phi}
\begin{cases}
\Gamma(\varphi)+ W(J)= J \varphi~~,\\
\phantom{x}  \\
\varphi = \dfrac{\partial}{\partial J} W(J)~~,\\
\phantom{x}  \\
J=\dfrac{\partial}{\partial \varphi} \Gamma(\varphi)~~.
\end{cases}
\end{equation}
The advantage is that, in a diagrammatic theory, the function $\Gamma(\varphi)$ is the generating function of One-Particle Irreducible (1PI) vertices. This means that, roughly speaking, in this approach all terms in the expansion in powers of $\Delta/N$ represented by one-particle reducible diagrams are summed up. Such terms represent an infinite series, thereby leading to a more accurate approximation \cite{zj}.

The effective potential can be expressed as
\begin{equation}
\label{eqn:eff_act}
	\Gamma(\varphi)=S(\varphi)+\frac{1}{2N} \ln D^{-1}(\varphi) + \Gamma_1(\varphi) + C
\end{equation}
where $-\Gamma_1(\varphi)$ is the sum of all 1PI vacuum diagrams of a new theory with action
\begin{gather}
     S(\phi; \varphi)=\frac{1}{2}D^{-1}(\varphi)\,\phi^2 + i \frac{g_3}{3!}\phi^3~~,\\
     D^{-1}(\varphi)= \Delta+ig_3\varphi~~.
\end{gather}
The (constant) term  $C = -\frac{1}{2N} \ln \left(\frac{2\pi}{N \mathcal{A}^2 }\right)$ appearing in \eqref{eqn:eff_act} ensures that $\Gamma(\overline{\varphi}) = 0$, where $\overline{\varphi} = \varphi(J=0)$, following from the normalization $W(J=0)=0$.  In practice, it merely fixes the value of ${\cal A}$ to
\begin{equation}
\label{eq:Norm_A}
  {\cal A} = \sqrt{\frac{2\pi}{N D^{-1}(\overline{\varphi})}}\, 
          e^{-N\bigl[S(\overline{\varphi}) + \Gamma_1(\overline{\varphi})\bigr]}~~,
\end{equation}
and will henceforth be omitted. In turn, the function $\Gamma_1(\varphi)$ can be written as a power series of $D(\varphi)/N$ (`loop expansion') starting with  an $O(1/N^2)$ term. Hence, to the leading $O(1)$ order, $\Gamma(\varphi)$ reads
\begin{equation}
\label{yyy}
   \Gamma(\varphi)\simeq S(\varphi)=  \frac{1}{2} \Delta\varphi^2+i \frac{g_3}{3!}\varphi^3~~.
\end{equation}

In order to compute $W(J)$ as $W(J) = J\varphi - \Gamma(\varphi)$ we have to eliminate $\varphi$ as function of $J$ using
\begin{equation}
\label{eqn:j_eq}
     J\equiv\frac{\partial}{\partial \varphi} \Gamma(\varphi)= \Delta\varphi +i \frac{g_3}{2} \varphi^2~~.
\end{equation}
Solving for $\varphi$ one finds
\begin{equation}
\label{eq:varphi_0}
\varphi=\varphi_0\equiv\frac{i}{\Delta^{-1} g_3}\left[1 - \sqrt{1+2ig_3\Delta^{-2}J}\right]~~,
\end{equation}
where the solution with $\partial^2 \Gamma(\varphi)/\partial\varphi^2 >0$ has been
taken \cite{Note_G2}.
Since $\overline{\varphi} = 0$, from \eqref{eq:Norm_A} it follows that
\begin{equation}
 \mathcal{A}=\sqrt{\frac{2\pi}{N\Delta}}~~.
\end{equation} 
Finally, starting from \eqref{eqn:def_phi}, straightforward algebra leads to \red{}
\begin{multline}
\label{dab}
%\begin{split}
   W(J)  = \frac{1}{3}\frac{iJ\Delta}{g_3} +\left[\frac{2}{3}\frac{iJ\Delta}{g_3}+\frac{1}{3}\frac{\Delta^3}{g_3^2}\right]\times\\
   \times\left[1-\sqrt{1+2g_3\Delta^{-2}iJ}\right]~~.
%\end{split}
\end{multline}

Summing up, from the second line of \eqref{eqn:P_lambda1}, one  obtains
\begin{equation}\label{cubic1}
  P_\lambda(\lambda;x) = \sqrt{\frac{N}{2\pi\Delta}} \, e^{-NF_0(\lambda)}~~,
\end{equation}
with
\begin{multline}
\label{eq:F_0}
%\begin{split}
 F_0(\lambda) = \frac{\Delta}{3 g_3}\Bigg\{(\mu-\lambda)
        +\left[2(\mu-\lambda)
             -\frac{\Delta^2}{g_3}\right]\times\\
             \times\left[1-\sqrt{1-2g_3\Delta^{-2}(\mu-\lambda)}\right]\Bigg\}~~.
% \end{split}
\end{multline}
\red{This approximation is valid provided 
\begin{equation}\label{root}
\mu-\lambda \leq \frac{\Delta^2}{2g_3}~~. 
\end{equation}}
By comparison, the Gaussian approximation holds as long as $|\lambda - \mu| = O(\sqrt{\Delta/N})$.

\subsubsection{Cubic approximation: next-to-leading order}

The $O(1/N)$ term of the cubic theory can be computed by including, in the approximate expression for $\Gamma(\varphi)$, the second term on the r.h.s of \eqref{eqn:eff_act}. This yields
\begin{multline}
   \Gamma(\varphi)\simeq S(\varphi)+\frac{1}{2N} \ln D^{-1}(\varphi)\\=  \frac{1}{2} \Delta\varphi^2+i \frac{g_3}{3!}\varphi^3
           +\frac{1}{2N} \ln \bigl[\Delta+ig_3\varphi\bigr]~~.
\end{multline}
The equation for  $\varphi$, which now reads
\begin{equation}
 \label{eqn:j_eq_2}
    J\equiv\frac{\partial}{\partial \varphi} \Gamma(\varphi)= \Delta\varphi +i \frac{g_3}{2} \varphi^2+ \frac{1}
     {2N}\frac{ig_3}{\Delta+ig_3\varphi}~~,
\end{equation}
can no longer be solved in closed form for $\varphi$. Nevertheless, by solving it numerically for $\varphi$, $W(J)$ can be evaluated for any $J$ directly as $W(J) = J\varphi - \Gamma(\varphi)$.

In our case, as $J=i(\mu -\lambda)$ is purely imaginary, so is the solution of \eqref{eqn:j_eq_2}. Therefore, redefining $i\varphi \to \varphi$, we have
\begin{equation}
\label{eq:Px_31}
  P_\lambda(\lambda;x) = {\cal C} \, e^{-NF(\lambda)}~~,
\end{equation}
with 
\begin{equation}
\label{eqn:F_3}
 \begin{split}
  F(\lambda) = 
	\frac{1}{2}\Delta \varphi^2 + \frac{g_3}{3} \varphi^3 
	    - \frac{1}{2N}&\frac{g_3 \varphi}{\Delta+g_3\varphi} \\
	&+\frac{1}{2N} \ln [\Delta+ g_3\varphi]~~,
\end{split}
\end{equation}
and where $\varphi \equiv \varphi(\lambda)$ denotes the solution of
\begin{equation}
 \label{eq:phi}
    \lambda - \mu =\Delta\varphi + \frac{g_3}{2} \varphi^2
        - \frac{1}{2N}\frac{g_3}{\Delta+g_3\varphi}~~.
\end{equation}
The normalization constant $\mathcal{C}$ is given by
\begin{equation}
  {\cal C} = \sqrt{\frac{N}{2\pi\bigl[\Delta + g_3\overline{\varphi}\bigr]}} \, 
       \exp\left\{N\left[ \frac{\Delta}{2}\overline{\varphi}^2  + \frac{g_3}{3!} \overline{\varphi}^3\right]\right\}~~,
\end{equation}
where $\overline{\varphi} \equiv \varphi(\mu)$. Now the probability distribution function $P_\lambda(\lambda;x)$ can be evaluated for any $\lambda$ by solving \eqref{eq:phi} numerically upon varying $\lambda$. \red{Note that this approximation holds if 
\begin{equation}\label{root2}
\Delta + g_3 \varphi > 0~~. 
\end{equation}}

\subsection{Probability distribution $P_I(I^\star)$}
\label{sub:P_I}

The maximal mutual information $I^\star = \log_2 Z$ depends on the quantity 
\begin{equation}
 y \equiv\frac{Z}{\sqrt{N}} = \int_{x_{\min}}^{x_{\max}} dx\, \sqrt{\lambda(x)}~~,
\end{equation}
where $\lambda(x)=\frac{1}{N}\sum_{i=1}^N f_i(x)$, cf. \eqref{zstar2}. Hence, in order to compute the full probability density $P_{\rm I}(I^\star)$ of $I^\star$, one has to evaluate the probability density of $y$,  given by
\begin{equation}
\begin{split}
\label{eq:P_y_1}
 P_{\rm y}(y) &=  \left\langle \delta \Bigl(y - \int dx\, \sqrt{\lambda}\Bigr)\right\rangle_{\lambda}
             \\
             & =
             \left\langle \int \frac{d\hat{y}}{2\pi}
               e^{-i\hat{y} \bigl(y - \int dx\, \sqrt{\lambda}\bigr)}
               \right\rangle_{\lambda}
             \\
             & =
             \int \frac{d\hat{y}}{2\pi} e^{-i\hat{y} y} 
                    \left\langle e^{i\hat{y} \int dx\, \sqrt{\lambda}}\right\rangle_{\lambda}~~,
 \end{split}
\end{equation}
where the subscript ``$\lambda$''  indicates the average over the function $\lambda(x)$ (functional average) with probability density functional ${\cal P}[\lambda]$. Note that,  at variance with the calculation of the average capacity, the correlations of the $f_i$s at different values of $x$ are relevant to compute $P_\textrm{I}$ and cannot be neglected. As a consequence, we shall limit the calculation to the Gaussian approximation in which ${\cal P}[\lambda]$ is given by \eqref{eq:P_lam_gau}.

The simplest way of computing $P_{\rm y}(y)$ for large $N$ starts with the observation that, for $N\gg1$, the functional ${\cal P}[\lambda]$ is expected to be sharply peaked about its mean. Therefore, the average $\avg{\cdots}_\lambda$ of a generic functional ${\cal F}[\lambda]$ of $\lambda$ can be evaluated as
\begin{widetext}
\begin{equation}
%  \begin{split}
   \left\langle {\cal F}[\lambda]\right\rangle_\lambda \simeq 
  {\cal F}[\mu] 
   + \int dx\,   \left.\frac{\delta {\cal F}[\lambda]}{\delta \lambda(x)}\right|_{\mu} 
                 \left\langle \lambda(x) - \mu(x) \right\rangle_\lambda 
  + \frac{1}{2}
        \int dx\, dx'\,   \left.\frac{\delta^2 {\cal F}[\lambda]}{\delta \lambda(x)\delta \lambda(x')}\right|_{\mu} 
                 \left\langle [\lambda(x) - \mu(x)][\lambda(x') - \mu(x')] \right\rangle_\lambda 
                  + \dotsc.
%  \qquad (M\to\infty)~~.
%\end{split}  
\end{equation}
\end{widetext}
For our purposes (see \eqref{eq:P_y_1}), ${\cal F}[\lambda] = e^{i\hat{y} \int dx\, \sqrt{\lambda}}$, so that
 \begin{equation}
%  F[\lambda] &= e^{i\hat{y} \int dc\, \sqrt{\lambda}},  \\  
  \frac{\delta {\cal F}[\lambda]}{\delta \lambda(x)}
            = \frac{i\hat{y}}{2\sqrt{\lambda(x)}}\, e^{i\hat{y} \int dx\, \sqrt{\lambda}}~~, 
\end{equation}
and
\begin{multline}
  \frac{\delta^2 {\cal F}[\lambda]}{\delta \lambda(x)\delta \lambda(x')} 
          = -\Bigg[ \frac{i\hat{y}}{4\lambda(x)^{3/2}} \, \delta(x-x')+\\
                    + \frac{\hat{y}^2}{4\sqrt{\lambda(x)\, \lambda(x')}}\Bigg]\, e^{i\hat{y} \int dx\, 
                    \sqrt{\lambda}}~~.
 \end{multline}
On the other hand, a Gaussian $\mathcal{P}[\lambda]$ as in  \eqref{eq:P_lam_gau} implies
\begin{gather}
%\begin{split}
 \left\langle \lambda(x) - \mu(x) \right\rangle_\lambda  = 0~~,\\
 \left\langle [\lambda(x) - \mu(x)][\lambda(x') - \mu(x')] \right\rangle_\lambda =
    \Delta(x,x')/N~~.
%\end{split}
\end{gather}
Putting pieces together, one gets
\begin{equation}
%\begin{split}
  \left\langle e^{i\hat{y} \int dx\, \sqrt{\lambda}}\right\rangle_{\lambda} 
  %&\sim
%  e^{i\hat{y} \int dc\, \sqrt{\mu}} \Bigl[            1  - \frac{i\hat{y}}{8N}\int dc\, \frac{\Delta(c,c)}{\mu(c)^{3/2}}
%           \\
%           &\phantom{============}
%           - \frac{\hat{y}^2}{8N}\int dc\, dc'\, \frac{\Delta(c,c')}{\sqrt{\mu(c)\, \mu(c')}}
%           + O(1/N^2)\Bigr]
%  \\
%  &
\simeq
  \exp\Bigl[ i\hat{y}\, \bar{y} - \frac{\Delta_z}{2N}\hat{y}^2 + O(1/N^2)\Bigr]~~,
%  \qquad N\to\infty,
%  \end{split}
\end{equation}
where
%\begin{align}
\begin{gather}
 \bar{y} = \int dx\,  \left[  \mu(x)^{1/2} - \frac{\Delta(x,x)}{8N\mu(x)^{3/2}}\right]~~\label{y_bar},
 \\
 \Delta_z = \frac{1}{4} \int dx\, dx'\,  \frac{\Delta(x,x')}{\sqrt{\mu(x)\, \mu(x')}}~~\label{deltaz}.
%\end{align}
\end{gather}
Substituting this expression back into \eqref{eq:P_y_1} and integrating over $\hat y$ we obtain
\begin{equation}
%\begin{split}
  P_{\rm y}(y) \simeq
  %&\sim \int \frac{d\hat{y}}{2\pi}           \exp\left[-\frac{\Delta_z}{2N}\hat{y}^2 -i\hat{y}\bigl(y - \bar{y}\bigr) + O(1/N^2)\right] \\
 %         &\sim
          \sqrt{\frac{N}{2\pi\Delta_z}}\,
     \exp\left[-\frac{N}{2\Delta_z}\bigl(y - \bar{y}\bigr)^2\right]     + O(1/N^2)~~.
%     \qquad N\to\infty
%\end{split}
\end{equation}
Because $Z = \sqrt{N} y$, see Eq. (\ref{zstar2}), we have
\begin{equation}
  P_{\rm Z}(Z) = \frac{dy}{dZ}\, \left.P_\textrm{y}(y)\right|_{y= \frac{Z}{\sqrt{N}}}
    =  \frac{1}{\sqrt{N}}\, \left.P_\textrm{y}(y)\right|_{y= \frac{Z}{\sqrt{N}}}~~,
\end{equation}
which immediately yields
\begin{equation}
  P_{\rm Z}(Z)\simeq \frac{1}{\sqrt{2\pi\Delta_z}}\,
     \exp\left[-\frac{1}{2\Delta_z}\Bigl(Z - \sqrt{N} \,\bar{y}\Bigr)^2\right]~~.
\end{equation}
Finally, from
\begin{equation}
  P_{\rm I}(I^\star) = \frac{dZ}{dI^\star}\, \left.P_{\rm Z}(Z)\right|_{Z= 2^{I^\star}}~~,
\end{equation}
we arrive at
\begin{equation}
\label{P_Istar}
  P_{\rm I}(I^\star)\simeq \frac{2^{I^\star}}{\sqrt{2\pi\Delta_z}}\, 
     \exp\left[-\frac{1}{2\Delta_z}\Bigl(2^{I^\star} - \sqrt{N} \bar{y}\Bigr)^2\right]~~,
\end{equation}
which represents our final expression of $P_\textrm{I}$ under the Gaussian approximation. Notice that, in absence of further constraints, the quantity $I^\star$ appearing here varies in principle from $-\infty$ to $+\infty$. \red{Hence, in this framework we cannot obtain an analytical estimate of the maximum achievable value of $I^\star$, namely $I^\star_{\rm opt}$.}

%\subsection{\red{Altro?}}

\section{Numerical results}

Following e.g. \cite{pre2}, we now specify the functions $\bar g_i$ and $\sigma_i^2$ to expressions \eqref{gbar} and  \eqref{sigma2} respectively. With this choice, $Z$ (Eq. \eqref{zeta}) can be written as
\begin{equation}
	Z= \sqrt{\frac{M_{\max}}{2\pi e}} \,\widetilde{Z}
\end{equation}
where 
\begin{equation}
\label{ztilde}
\widetilde{Z}=\int_{x_{\min}}^{x_{\max}} \left[\sum_{i=1}^N\frac{h_i^2 K_i^{2h_i}}{x\left[h_i^2 K_i^{2h_i}+x^{1-h_i}\left(x^{h_i}+K_i^{h_i}\right)^3\right]}\right]^{1/2}dx~~.
\end{equation}
In turn, the maximal mutual information reads
\begin{equation}
\label{Itilde}
	I^{\star}\equiv \log_2 Z= \frac{1}{2}\log_2 \frac{M_{\max}}{2\pi e} + \widetilde{I}^{\star},
\end{equation} 
where $\widetilde{I}^{\star}=\log_2 \widetilde{Z}$ is independent of $M_{\max}$. For this reason, we shall henceforth focus our attention on $\widetilde I^\star$ as opposed to $I^{\star}$.

\begin{figure*}
	\includegraphics[width=\textwidth]{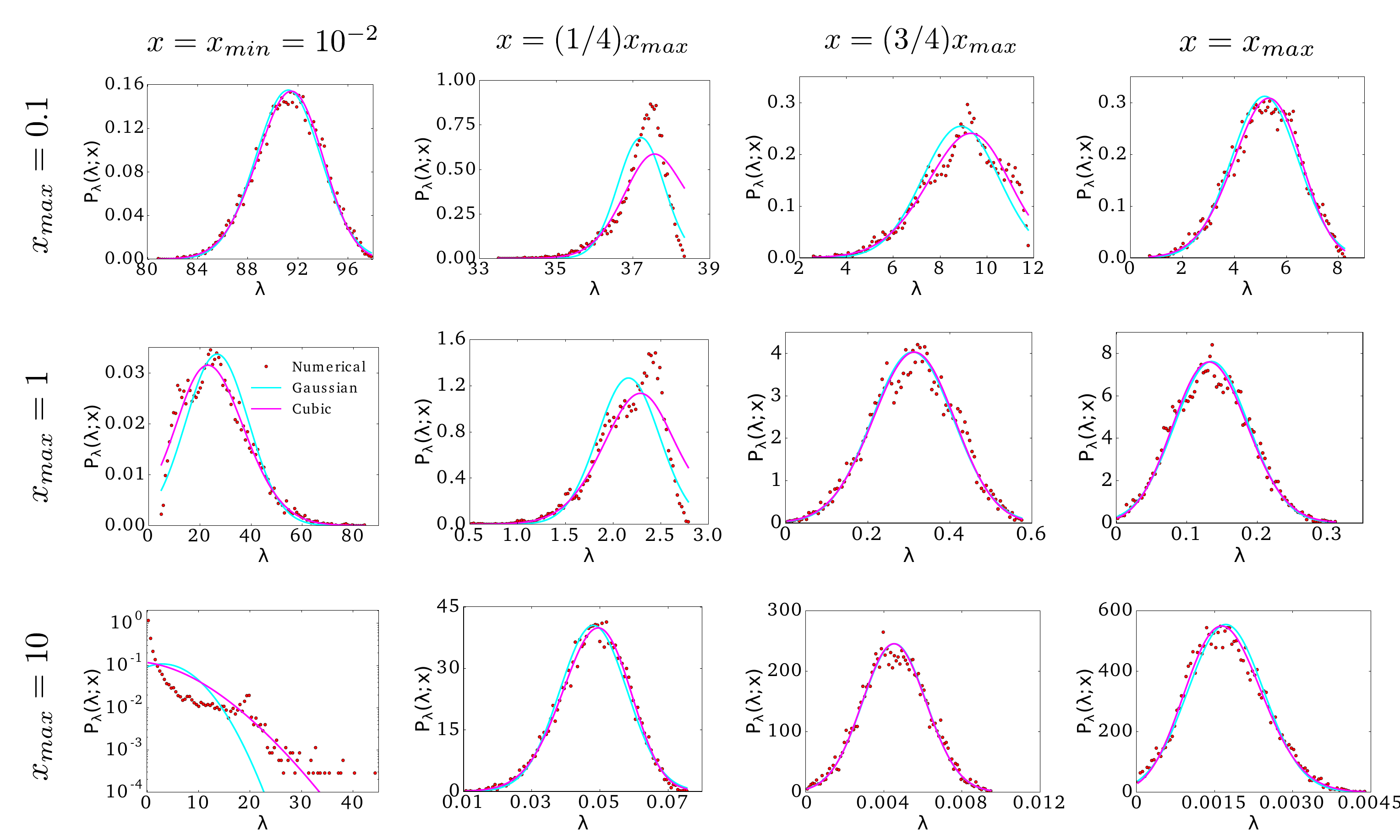}
	\caption{Probability distributions $P_\lambda(\lambda;x)$ at four different values of the \red{controller} level $x$ for $N=5$ targets. The vector $\boldsymbol{K}$ was sampled from a uniform distribution in $[x_{\min},x_{\max}]$, $x_{\min}=10^{-2}$, $x_{\max}=0.1,1,10$ (respectively top, middle and bottom row). Hill indices were fixed to $h_i=2$, $\forall i=1,\dots,N$. Red markers represent the numerical results, while solid lines display analytical results corresponding to the Gaussian (Eq. \eqref{eqn:P_lambda_2}, cyan) \red{and cubic (Eqs. \eqref{eq:Px_31}-\eqref{eq:phi}, magenta) approximations.}} %\red{Note that, while the Gaussian estimate is defined for all the values of $\lambda$, cubic 1 and cubic 2 are only defined for values of $\lambda$ such that conditions (\ref{root}) and, respectively, (\ref{root2}) are satisfied.}}
	\label{fig:FIG3}	
\end{figure*}

In order to evaluate $\avgs{\widetilde{I}^{\star}}_{\rm ann}=\log_2 \avgs{\widetilde{Z}}$, we have computed $\avgs{\widetilde Z}$ using the different approximations discussed for $P_\lambda(\lambda;x)$, namely the Gaussian estimate \eqref{eqn:P_lambda_2} \red{and the cubic estimate \eqref{eq:Px_31}-\eqref{eq:phi}}. As these quantities depend on the cumulants of the $f_i$s, we have first computed,  for each value of the input variable $x$ in the interval $[x_{\min},x_{\max}]$, the quantities $\mu$, $\Delta$ and $g_3$ defined in Eqs. \eqref{eq:mu_x}-\eqref{eq:gn_x}. To give an idea of the quality of the different approximation schemes, we showcase in Fig. \ref{fig:FIG3} the distributions $P_\lambda(\lambda;x)$ obtained by fixing the Hill indices $\boldsymbol{h}$ and sampling $\boldsymbol{K}$ uniformly in $[x_{\min},x_{\max}]$, for $x_{\max}=0.1$ (top row), $x_{\max}=1$ (middle row) and $x_{\max}=10$ (bottom row) and at the values of $x$ corresponding to the different columns. For $x_{\max}=0.1$ and $x_{\max}=1$ the Gaussian distribution appears to provide an accurate approximation to the true $P_\lambda(\lambda;x)$ at sufficiently large values of $x$, while corrections to the Gaussian picture are necessary to describe the tail of $P_\lambda(\lambda;x)$ for smaller values of $x$, especially for $x=x_{\min}$ and $x_{\max}\gg x_{\min}$ (although even the cubic theory falls short of describing the small-$\lambda$ regime in that case). %\red{While the Gaussian estimate is defined for all the values of $\lambda$, the leading-order and next-to-leading order cubic estimates are defined respectively for $\mu-\lambda \leq \frac{\Delta^2}{2g_3}$ and $\mu-\lambda < \frac{\Delta^2}{2g_3}$. This explains why the curves representing the cubic estimates in Fig. \ref{fig:FIG3} do not cover the whole range of $\lambda$ numerical values. Moreover, since the next-to-leading-order cubic estimate contains a logarithm (Eq. \eqref{F1}), when $\mu-\lambda \simeq \frac{\Delta^2}{2g_3}$ $P_\lambda(\lambda;x)$ explodes.}

\begin{figure*}
	\includegraphics[width=\textwidth]{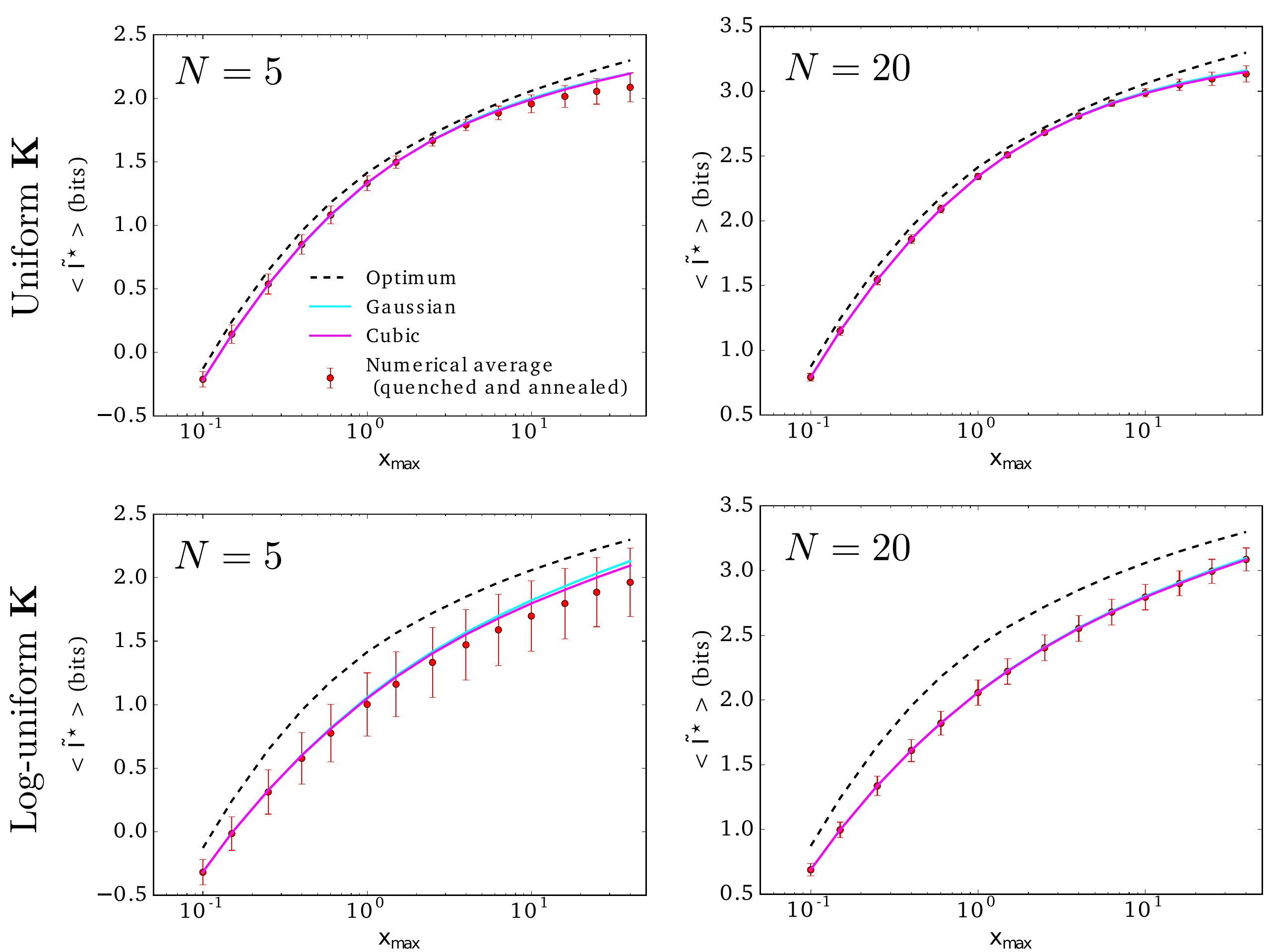}
	\caption{Comparison between numerical results for the average capacity $\avgs{\log_2 \widetilde{Z}}$ (markers) and analytical results for the annealed average $\log_2\avgs{\widetilde{Z}}$ obtained within the Gaussian (Eq. \eqref{eqn:P_lambda_2}, cyan) \red{and cubic (Eqs. \eqref{eq:Px_31}-\eqref{eq:phi}, magenta) approximations. Vectors $\boldsymbol{K}$ in the top panels and in the bottom panels are sampled respectively from a uniform and a uniform in log-scale distributions in $[x_{\min},x_{\max}]$}, $x_{\min}=10^{-2}$, while Hill indices are fixed to $h_i=2$, $\forall i=1,\dots,N$. The dashed line gives the value of $\widetilde{I}^\star_{\rm \opt}$ for each $x_{\max}$. \red{The absolute difference between quenched and annealed averages estimated at $x_{\max}=40$ through the argument given in Appendix B in the uniform (log-uniform) case is $\simeq 4 \cdot 10^{-3}$ ($2 \cdot 10^{-2}$) for $N=5$ and $\simeq 10^{-3}$ ($3 \cdot 10^{-3}$) for $N=20$. Note that in the log-uniform case, for $N=20$ and large $x_{\max}$, the difference between the cubic approximation and the quenched numerical average is comparable to the difference between quenched and annealed numerical averages.}}
	\label{fig:FIG4}	
\end{figure*}

Fig. \ref{fig:FIG4} shows how numerical results for $\avgs{\widetilde{I}^{\star}}=\avgs{\log_2 \widetilde{Z}}$ compare with the annealed approximation $\avgs{\widetilde{I^\star}}_{\rm ann}$ as a function of $x_{\max}$ (and fixed $x_{\min}$) for $N=5$ and $N=20$. \red{In the case of fixed $\boldsymbol{h}$ and $\boldsymbol{K}$ uniformly sampled in $[x_{\min},x_{\max}]$ (top panels),} one sees that already for $N=5$ the Gaussian approximation describes numerical results with good accuracy, as slight deviations only occur at large $x_{\max}$. \red{The cubic approximation slightly improves the Gaussian picture.}
Notice that the ensemble average systematically trails the optimum by a \red{few per cent. When the distribution of $K$ is uniform in log-scale (see bottom panels), the qualitative scenario is unchanged although the relative difference between the mean and the optimum increases (for $N=20$, it generically ranges between 10 and 20\%). In this case the improvement provided by the cubic approximation is more appreciable, especially for $N=5$ and large $x_{\max}$.}

\begin{figure*}
	\includegraphics[width=\textwidth]{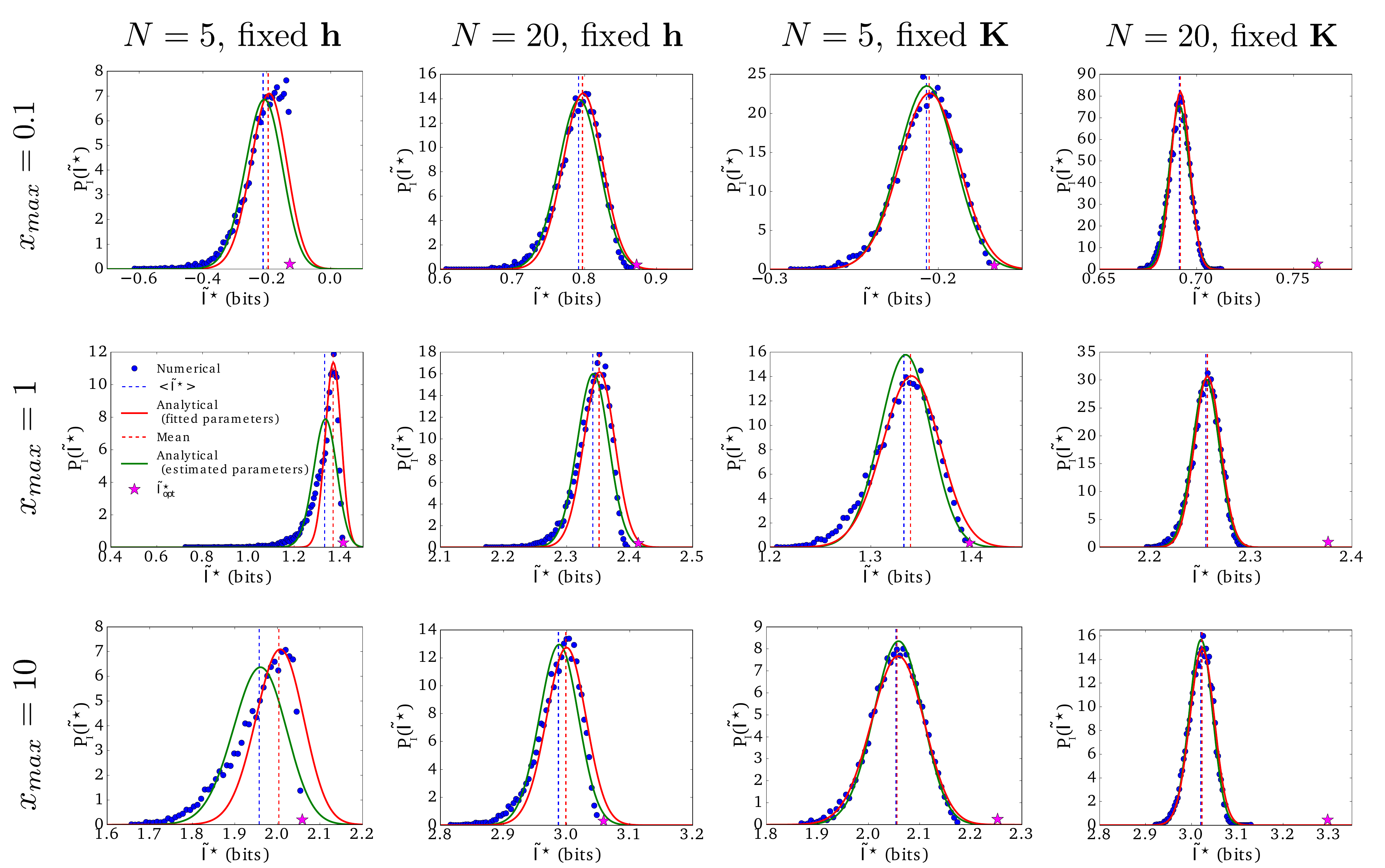}
	\caption{Probability distributions $P_{\rm I}(\widetilde{I}^\star)$ for different $x_{\max}$ (rows) and $N$ (columns). Markers represent the distribution obtained numerically from $10^4$ samplings of $\boldsymbol{K}$ from a uniform distribution in $[x_{\min},x_{\max}]$ ($x_{\min}=10^{-2}$) and Hill indices fixed at $h_i=2$ $\forall i=1,\dots,N$ (left two columns), and from $10^4$ samplings of $\boldsymbol{h}$ from a uniform distribution in $[1,5]$ and fixed $\boldsymbol{K}$ (chosen randomly in $[0.01,0.1]$ for the case $x_{\max}=0.1$ and properly re-scaled for the cases $x_{\max}=1$ and $x_{\max}=10$). The red and green lines represent the analytical form \eqref{P_Istar} with fitted and estimated parameters, respectively (see text for details). In all panels, purple stars denote the value of $\widetilde{I}^\star_{\opt}$.}
	\label{fig:FIG5}
\end{figure*}

Fig. \ref{fig:FIG5} displays how the distribution of capacities $P_\textrm{I}(\widetilde{I}^\star)$ obtained numerically compares with the analytical expression \eqref{P_Istar}, for two ensembles of regulatory motifs: one with fixed Hill indices $\boldsymbol{h}$ and quenched random $\boldsymbol{K}$, and the other with fixed $\boldsymbol{K}$ and quenched random $\boldsymbol{h}$. Specifically, we have evaluated \eqref{P_Istar} in two ways: first, by estimating its parameters $\bar y$ and $\Delta_z$ (Eq.s \eqref{y_bar} and \eqref{deltaz}) from numerical data (green curves); second, by straightforwardly fitting them to data (red curves). 
In the fixed-$\boldsymbol{h}$ ensemble, estimated parameters lead to a good agreement between analytic expression and numerical results only for larger $N$, while fitted distributions appear to provide a slightly better description of the data already for $N=5$. In the fixed-$\boldsymbol{K}$ ensemble, the agreement is generically good already for $N=5$ for both estimated and fitted parameters. 

\begin{figure}
	\includegraphics[width=0.5\textwidth]{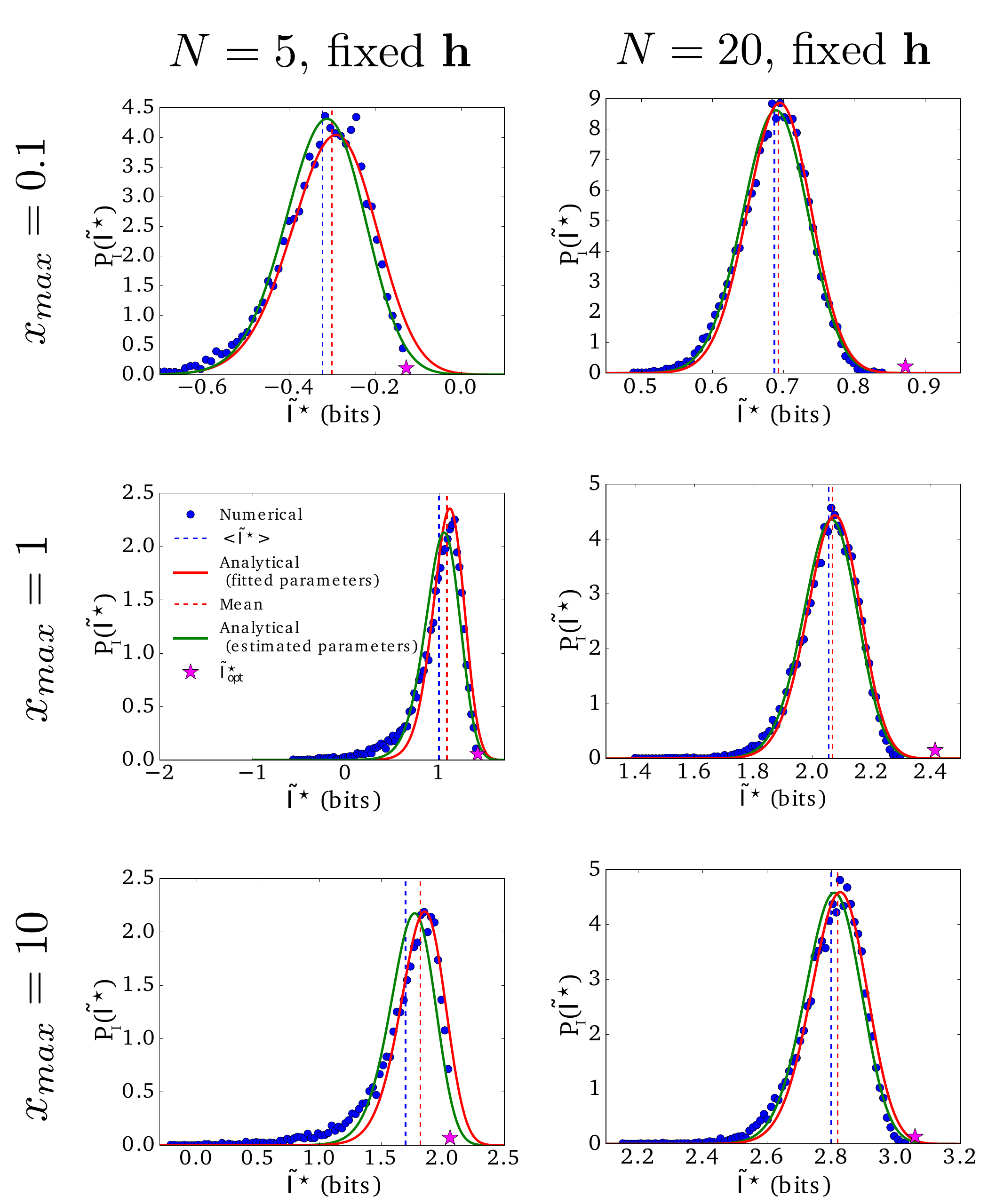}
	\caption{\red{Same as Fig. \ref{fig:FIG5} but with $\boldsymbol{K}$ sampled from a uniform distribution in log-scale in $[x_{\min},x_{\max}]$, with fixed $x_{\min}=10^{-2}$ and $h_i=2$ $\forall i=1,\dots,N$.}}
	\label{fig:FIG6}
\end{figure}

\red{Finally, Fig. \ref{fig:FIG6} shows the distribution of capacities obtained with fixed Hill indices $\boldsymbol{h}$ and quenched random $\boldsymbol{K}$ sampled from a uniform distribution in log-scale in $[x_{\min},x_{\max}]$. Distributions appear to be broader than in the previous case, while estimated parameters provide a good agreement with numerical results already for $N=5$. The qualitative outlook is however similar to that obtained for uniform $\boldsymbol{K}$.}

%\section{Multiple solutions?}

%\red{Open question}

\section{Discussion}

Over the past few years, field-theoretic techniques have found novel application grounds in the quantitative study of signal processing in molecular networks (see e.g. \cite{thir} for a recent example). In this paper, we have used approximation methods routinely employed in statistical field theory to characterize states of optimal (static) information flow in elementary regulatory motifs in which a single variable $x$ controls $N$ output variables $g_i$ ($i=1,\ldots,N$). 

We have specifically focused on the ensemble of motifs generated by randomly sampling kinetic parameters according to a prescribed distribution. Our analysis started from the observation that optimal properties are determined by the statistics of $\lambda(x)$, Eq. \eqref{lambda}, at fixed $x$. As this quantity involves a sum of the $N$ functions $f_i(x)$ defined in \eqref{accai}, one may hope to apply the Central Limit Theorem and describe its large-$N$ statistics via a Gaussian approximation. The latter amounts to truncating the exact expression for the action $S$ (Eq. \eqref{eq:S_x}) that defines the probability density $P_\lambda(\lambda;x)$ (Eq. \eqref{eqn:P_lambda1}) to the first term, and leads to expression \eqref{eqn:P_lambda_2} for $P_\lambda(\lambda;x)$. In turn, it provides, via the annealed approximation, the Gaussian estimates for the mean maximum mutual information $\avg{I^\star}_{\rm ann}$ and for the probability density $P_\textrm{I}(I^\star)$. \red{We have also derived expressions for $P_\lambda(\lambda;x)$ beyond the Gaussian approximation, by including the second term (third-order) in $S$, obtaining expression \eqref{cubic1} at the leading order, and \eqref{eq:Px_31}-\eqref{eq:phi} at the next-to-leading order.} %(we also derived the analytical expression \eqref{cubic2} in the limit $N \gg 1$). 

Our results provide very good estimates for $\avg{I^\star}$ (see Fig. \ref{fig:FIG4}) even for $N$ as small as 5. On the other hand, the estimate of $P_\textrm{I}$ obtained from the Gaussian approximation appears to be generically accurate for large enough $N$. For smaller $N$ (e.g. $N= 5$), the precise form of the ``disorder distribution'' $Q$ appears to be important.

The major limitation of the theory we presented lies, in our view, in the assumption that no direct inter-target interactions occur. It is not hard to understand that, while allowing for a richer phenomenology (see e.g. \cite{pre3}), the presence of correlations between target levels considerably complicates calculations. Advancing the theory in that direction would greatly broaden our grasp of how the effectiveness of regulatory circuits is modulated by topological and/or kinetic heterogeneities. 

While working in the large-$N$ limit may seem to be unrealistic, real regulatory modules often involve molecular species controlling large numbers of targets. An example is provided by non-coding regulatory RNAs like miRNAs, some of which are known to regulate hundreds of RNA species in eukaryotes. In this light, our results highlight some features with potential biological significance. Firstly, assuming optimal input control and strong randomness in parameters, the mean information flow (i.e. the average over parameters of the maximum mutual information achievable between input and outputs) always appears to be very close to the optimum (generically, within a few \red{per cent} for $N$ between 5 and 20). This suggests that, in the presence of multiple targets, where information can potentially be exchanged across many channels by exploiting kinetic heterogeneities, optimizing kinetic parameters \red{might only provide} relatively minor and possibly very costly improvements to noise processing. \red{Albeit roughly, such a scenario might indeed explain why, in certain transcriptional \cite{gerland} and post-transcriptional \cite{clash} regulatory systems, a substantial fraction of input-output couplings appear to be sub-optimal or even non-specific.}  

On the other hand, achieving optimal control of the input variable appears to be crucial if information flow is to be optimized (or nearly optimized). \red{Indeed, in the small noise limit, the information flow $I$ due to a sub-optimal input distribution undershoots the optimal value $I^\star$ by the KL-divergence between the sub-optimal ($P_{\rm x}$) and the optimal ($P_{\rm x}^\star$) input distributions, i.e.
\begin{equation}
I=I^\star-D_{\rm KL}(P_{\rm x}||P_{\rm x}^\star)~~.
\end{equation}
Numerically, sub-optimal inputs have been shown to cause significant losses in the efficiency of information flow when capacities are sufficiently large, specifically larger than about 1 bit \cite{pre1}. By contrast, we find here that the losses induced by sub-optimal parameter values are very modest.} This suggests that, when the input variable is endogenously controlled, regulatory systems may want to invest resources into fine-tuning its distribution. In turn, a potentially important cost-benefit trade-off may arise in optimizing the input variable. Such a trade-off is likely to be especially limiting when the noise affecting the input-output channel can not be considered ``small''. \red{It would therefore be especially interesting to characterize these aspects more precisely by extending the scenario described here beyond the small noise approximation, e.g. along the lines discussed in \cite{pre1}.}

\section*{Appendix A. The Small Noise Approximation}

For sakes of completeness, we recapitulate here the optimal information flow scenario in the small noise approximation, focusing for simplicity on the case $N=1$ (the extension to generic $N$ is straightforward). We assume a Gaussian input-output channel $P(g|x)$, with mean $\bar g(x)$ (taken to be positive and invertible) and variance $\sigma^2_g(x)$. Assuming that the latter quantity is ``small'', $P(g|x)$ behaves roughly as a $\delta$-distribution under integration. This implies that the output distribution $P_{\out}(g)$ is approximately given by 
\begin{equation}
P_{\out}(g)\equiv\int P(g|x)P_\inn(x)dx\simeq %\int\delta[g-\bar g(x)]P_\inn(x)dx\\
%\simeq\frac{1}{|\bar g'(\ovl{x})|}\int\delta(x-\ovl{x})P_\inn(x)dx=
\frac{P_\inn(\ovl{x})}{|\bar g'(\ovl{x})|}~~,
\end{equation}
where $\ovl{x}\equiv \bar g^{-1}(g)$. In turn, because $\bar g(\ovl{x})=g$, we have
\begin{equation}
[g-\bar g(x)]^2\simeq [\bar g'(\ovl{x})]^2 (x-\ovl{x})^2~~,
\end{equation}
so that, from Bayes' rule,
\begin{multline}
P(x|g)\simeq \frac{P_\inn(x)}{P_\inn(\ovl{x})}\sqrt{\frac{[\bar g'(\ovl{x})]^2}{2\pi\sigma^2_g(x)}}\,\,\, e^{-\frac{1}{2}\frac{[\bar g'(\ovl{x})]^2}{\sigma^2_g(x)}(x-\ovl{x})^2}\\
\equiv \frac{P_\inn(x)}{P_\inn(\ovl{x})}\,\,G[\bar g^{-1}(g),\sigma^2_x(g)]~~,
\end{multline}
where $G[A,B]$ denotes a Gaussian distribution with mean $A$ and variance $B$, and $\sigma^2_x(g)\simeq\sigma^2_g(x)/[\bar g'(\ovl{x})]^2$. 
%\begin{equation}
%\sigma^2_x(g)\simeq\frac{\sigma^2_g(x)}{[f'(\ovl{x})]^2}
%\end{equation}
%Note that
%\begin{equation}
%\int P(x|g)dx\simeq\frac{1}{P_\inn(\ovl{x})}\int \delta(x-\ovl{x})P_\inn(x)dx=1~~.
%\end{equation}

Summing up, ``small'' $\sigma^2_g(x)$ implies
\begin{gather}
P_{\out}(g)\simeq\frac{P_\inn[\bar g^{-1}(g)]}{|\bar g'[\bar g^{-1}(g)]|}~~,\\
P(x|g)\simeq\frac{P_\inn(x)}{P_\inn[\bar g^{-1}(g)]}\,\,G[\bar g^{-1}(g),\sigma^2_x(g)]~~.
\end{gather}
%The key question is: can we estimate how good this is in terms of information flow (\underline{without optimizing} over the input distribution)?
With our choice for $P(g|x)$, the mutual information takes the form
\begin{multline}
I(g;x)=-\int dx P_\inn(x)\log_2 \sqrt{2\pi e \sigma^2_g(x)}\\-\int dx P_\inn(x)\int dg P(g|x)\log_2 P_\out(g)~~.
\end{multline}
However, again approximating $P(g|x)$ with a Dirac-$\delta$ under integration, we have
\begin{equation}
\int dg P(g|x)\log_2 P_\out(g)\simeq\log_2 P_\out[\bar g(x)]~~,
\end{equation}
where $P_\out[\bar g(x)]\simeq P_\inn(x)/|\bar g'(x)|$ (since $\bar g^{-1}(\bar g(x))=x$). Hence
\begin{equation}\label{I_sna}
I(g;x)\simeq S[P_\inn(x)]-\int dx P_\inn(x)\log_2\sqrt{\frac{2\pi e \sigma^2_g(x)}{[\bar g'(x)]^2}}~~,
\end{equation}
where $S[P_\inn(x)]=-\int dx P_\inn(x)\log_2 P_\inn(x)$. By maximizing $I$ over $P_\inn$ (with an appropriate constraint enforcing normalization) one finds that the optimal input distribution is given by $P_\inn^\star(x)\propto |\bar g'(x)|/\sigma_g(x)$. This, together with expression \eqref{I_sna}, replicates the results obtained e.g. in \cite{pre2}.

\section*{Appendix B. On the difference between quenched and annealed averages}

Here we provide an estimate of the difference between the quenched average $\avg{\log_2 Z}$ and the annealed average $\log_2 \avg{Z}$. (Note that, by Jensen's inequality, $\log_2\avg{Z}\geq\avg{\log_2 Z}$.) Recalling that $I^\star\in[0,I^\star_{\rm opt}]$, $Z$ is a random variable defined in $[1,2^{I^\star_{\rm opt}}]$. We shall denote its probability distribution by $P(Z)$. The quenched average is given by
\begin{equation}
	\avg{\log_2 Z}= \int_{1}^{2^{I^\star_{\rm opt}}} P(Z) \log_2 Z \,dZ~~.
\end{equation}
Introducing the variable $t$ as $Z=\avg{Z}+t$, we can easily isolate the annealed average $\log_2 \avg{Z}$ to obtain
\begin{multline}
\avg{\log_2 Z}= \log_2 \avg{Z} +\\
+ \frac{1}{\ln 2}\int_{1-\avg{Z}}^{2^{I^\star_{\rm opt}}-\avg{Z}} P(t) \ln \left(1+\frac{t}{\avg{Z}}\right)\,dt~~,
\end{multline}
where $P(t) \equiv P(t+\avg{Z})$. The integral could in principle be evaluated using the Maclaurin series
\begin{equation}
	\ln(1+x)=-\sum_{n=1}^{\infty} \frac{(-x)^n}{n}~~,
\end{equation}
and integrating term-by-term. However, this is possible only if the series is convergent, i.e. in this case only if $|x|<1$. This translates into the constraint $2^{I^\star_{\rm opt}}<2\avg{Z}$ that, from numerical simulations, turns out to be always satisfied in our system. Therefore, the first non-zero term of the series (corresponding to $n=2)$ represents an estimate of the error made upon substituting $\avg{\log_2 Z}$ with $\log_2 \avg{Z}$. We obtain
\begin{equation}
	\avg{\log_2 Z}= \log_2 \avg{Z} - \frac{1}{2\ln 2}\frac{\sigma^2}{\avg{Z}^2} + O\left(\langle\left(Z-\avg{Z}\right)^3 \rangle\right)~~,
\end{equation} 
where 
\begin{equation}
	\sigma^2= \int_{1}^{2^{I^\star_{\rm opt}}} P(Z) \left(Z-\avg{Z}\right)^2\, dZ
\end{equation}
is the variance of the probability distribution $P(Z)$. 


\begin{thebibliography}{99}
\bibitem{Note_G2} The inverse of the second derivative of $\Gamma(\varphi)$ is equal to the second cumulant $\langle\phi^2\rangle_c$ in the theory described by the action $S(\phi)$ and must be positive.
\end{thebibliography}
\end{document}